\documentclass{article}
\usepackage{style/spconf,amsmath,graphicx}
\usepackage{enumitem}
\usepackage{adjustbox}
\usepackage[page]{appendix}
\usepackage[hyphens]{url}
\usepackage{hyperref}
\hypersetup{colorlinks=true}
\usepackage{listings}

\usepackage{amsfonts}
\usepackage{booktabs,multirow}
\usepackage{adjustbox}
\usepackage{algorithm}
\usepackage{algpseudocode}
\usepackage{color}
\usepackage{cite}
\usepackage{xcolor}

\newcommand{\mat}[1]{\boldsymbol{\mathbf{#1}}} %
\def\etal{{\textit{et al.}}}

\newcommand{\footnoteurl}[1]{\footnote{\scriptsize \url{#1}}}

\usepackage{caption}
\def\vspacereduce{{\vspace{-5pt}}}

\title{USM-SCD: Multilingual Speaker Change Detection Based on Large Pretrained Foundation Models}
\name{
\begin{tabular}{cc}
Guanlong Zhao, Yongqiang Wang, Jason Pelecanos\\Yu Zhang, Hank Liao, Yiling Huang, Han Lu, Quan Wang
\end{tabular}}
\address{Google LLC, USA \\
\texttt{\normalsize \{guanlongzhao,yqw,pelecanos,ngyuzh,hankliao,yilinghuang,luha,quanw\}@google.com}}
\begin{document}
\ninept
\maketitle
\begin{abstract}
\vspace{6pt}
We introduce a multilingual speaker change detection model (USM-SCD) that can simultaneously detect speaker turns and perform ASR for 96 languages. This model is adapted from a speech foundation model trained on a large quantity of supervised and unsupervised data, demonstrating the utility of fine-tuning from a large generic foundation model for a downstream task. We analyze the performance of this multilingual speaker change detection model through a series of ablation studies. We show that the USM-SCD model can achieve more than 75\% average speaker change detection F1 score across a test set that consists of data from 96 languages. On American English, the USM-SCD model can achieve an 85.8\% speaker change detection F1 score across various public and internal test sets, beating the previous monolingual baseline model by 21\% relative. We also show that we only need to fine-tune one-quarter of the trainable model parameters to achieve the best model performance. The USM-SCD model exhibits state-of-the-art ASR quality compared with a strong public ASR baseline, making it suitable to handle both tasks with negligible additional computational cost.
\end{abstract}
\vspacereduce
\section{Introduction}
\label{sec:intro}

Speaker change detection (SCD) \cite{ajmera2004robust} is the process of identifying the speaker turn points in a multi-speaker audio stream. SCD has broad applications in enhancing speaker diarization accuracy \cite{xia2022turn,wang2022highly}, improving Automatic Speech Recognition (ASR) quality \cite{sari2020auxiliary}, generating line breaks in captions to boost readability and accessibility \cite{donabauer2021making}, and augmenting textual prompts for multi-modal large language models (LLMs) \cite{tsimpoukelli2021multimodal}.

Conventionally, SCD is achieved by using a neural network to map acoustic features or speaker embeddings \cite{hruz2017convolutional,yin2018neural,aronowitz2020context} to a frame or segment level yes/no speaker change prediction. The neural network is generally trained by minimizing the binary cross entropy loss between the ground-truth SCD labels and the predictions. These conventional approaches have various limitations. First, they require accurate timing information of the speaker change point, which is difficult to obtain since marking speaker change timestamps is a highly subjective process for human annotators. Second, the methods that use purely acoustic information ignore rich semantic information in the audio. Third, the methods that use speaker embeddings utilize sensitive biometric information that can be exploited for unintended purposes and are sub-optimal from a privacy point of view~\cite{de2017europe}.

A few recent studies \cite{xia2022turn,zhao2023scd,wu2023speaker} explore using ASR-based approaches to detect word-level speaker changes to mitigate the aforementioned issues with conventional models. Xia \etal \cite{xia2022turn} propose an SCD model using a Transformer-Transducer (T-T). Specifically, they augment the text transcription of the spoken utterance with a special speaker turn token \texttt{<st>}, and then train the model to output both regular text tokens and the special speaker turn token. This model does not need accurate timestamps for training since the T-T model is trained in a seq2seq fashion and does not need forced-alignment to provide training targets. The model also utilizes both acoustic and linguistic information in the input audio. As a follow up of that work, in \cite{zhao2023scd} we propose a training loss that penalizes speaker change false acceptance and false rejection errors in the N-best hypotheses to further enhance performance. Wu \etal \cite{wu2023speaker} add an additional SCD module on top of an existing T-T ASR network to optimize the SCD and ASR tasks separately.

Recent advances in self-supervised learning have ushered in a new era for speech tasks. Large pretrained foundation models \cite{bommasani2021foundation} have led to significant performance improvement in various downstream speech tasks including emotion recognition \cite{pepino2021emotion}, language identification \cite{liu2022efficient}, voice activity detection \cite{kunevsova2023multitask}, and mispronunciation detection \cite{xu2021explore}. In this work we take advantage of the recent Google Universal Speech Model (USM) \cite{zhang2023usm} framework to build an SCD model that is capable of recognizing speaker changes in 96 languages. In addition, the performance of prior ASR-based models is limited by the quantity of supervised SCD data available for individual languages, leading to lower performance. We explore the benefit of using a large quantity of unsupervised and supervised multilingual ASR data for model pretraining. The major contributions of this paper include (1) a 96-language SCD model that significantly outperforms the previous monolingual baseline; (2) detailed ablation studies of the proposed multilingual SCD model.

\vspacereduce
\section{Method}
\label{sec:method}

First, we build a pretrained model as the foundation model. We then fine-tune the foundation model with data annotated with speaker changes.

\vspacereduce
\subsection{Backbone model}
\label{sec:method:backbone}

At a high level, the backbone model architecture used in this work consists of a Conformer encoder \cite{gulati2020conformer} and a Connectionist Temporal Classification (CTC) \cite{graves2006connectionist} decoder. The inputs are mel-spectra features and a one-hot vector representing the language of the utterance.

We pass the input features through mean variance normalization, SpecAugment \cite{park2019specaugment} (only for training), and multiple 2D-convolution layers (denoted as the \textit{feature encoder}) to reduce the input frame rate, similar to the setup in wav2vec 2.0 \cite{baevski2020wav2vec}. We then append the features with a one-hot language embedding. The concatenated features are then projected by a linear input projection layer to match the dimension of the Conformer encoder, which takes the projection layer outputs as its inputs. The Conformer encoder is trained with chunk-wise attention \cite{zhang2023usm}. The output of the Conformer encoder is passed to a linear projection layer, outputting logits that correspond to WordPiece tokens. The model is trained with the CTC loss. We do not use the RNN-T paradigm \cite{graves2012sequence} in this work due to its slow training speed as a result of its auto-regressive nature, which is especially prevalent when training large models with billions of parameters.

\vspacereduce
\subsection{Pretraining}

There are various pretraining techniques. In this work, we explore both supervised and unsupervised pretraining methods.

\vspacereduce
\subsubsection{BEST-RQ pretraining}

We select BEST-RQ \cite{chiu2022self} as the unsupervised method to pretrain our networks. BEST-RQ provides a simple framework with a small number of hyperparameters for unsupervised training on large-scale unlabeled audio data. BEST-RQ applies a random-projection quantizer to map speech signals to discrete labels to enable BERT-style pretraining for ASR encoders. The quantizer randomly initializes a matrix and a codebook, and uses the matrix to project the input speech signals and the codebook to find the nearest vector, where the index of the vector serves as the label. The pretraining process masks the speech signals and feeds them to the ASR encoder that learns to predict labels of the masked segment. The random projection performs dimension reduction for the speech signals while the random codebook provides an approximated discrete representation of the data distribution. Both the randomly initialized matrix and codebook are fixed during the pretraining process. In this study, the encoder in the BEST-RQ system employs the same model architecture as the Conformer encoder described in Sec. \ref{sec:method:backbone}.

\vspacereduce
\subsubsection{ASR pretraining}

For supervised pretraining, we initialize the Conformer encoder's weights from the BEST-RQ model's encoder and fine-tune it on ASR data to predict text from audio.

\vspacereduce
\subsection{SCD fine-tuning}

For the SCD task, we fine-tune the pretrained model with speaker change data, and we refer to this type of model as \textbf{USM-SCD}.

We warm start the backbone model's Conformer encoder from a pretrained model's encoder. The decoder projection layer is always randomly initialized. The training targets are WordPiece tokens augmented with speaker change annotations. To create training targets, we add a special speaker change token \texttt{<st>} between two different speakers' transcripts (e.g. ``hello how are you \texttt{<st>} I am good \texttt{<st>}'') to model speaker changes during training. Compared with audio-only SCD models~\cite{yin2018neural}, this model may more directly utilize the language semantics as a signal for speaker segmentation. For inference, we perform an ASR decoding with the SCD model, and identify the speaker change tokens. We use the timestamps of the predicted speaker turn tokens in the evaluation.

\vspacereduce
\subsection{Speaker change token posterior scaling}
\label{sec:method:scd_posterior}

The speaker change tokens are relatively scarce in the training data. To encourage the model to output speaker change tokens, we can apply a scaling factor to the posterior probability of the speaker change token $p(\texttt{<st>}|\mat{X})$ during decoding, where $\mat{X}$ is the model input. Assuming \textit{greedy} decoding (see section 3.2 of \cite{graves2006connectionist}) during inference, this can be achieved by multiplying $p(\texttt{<st>}|\mat{X})$ with a constant factor $\lambda > 1$, i.e., $p'(\texttt{<st>}|\mat{X}) = \lambda \cdot p(\texttt{<st>}|\mat{X})$. Effectively this increases the posterior probability of the \texttt{<st>} token. \textit{Greedy} decoding simplifies the process since we do not need to redistribute the rest of the probability mass as a result of the scaling. In practice, we operate on the log posterior probability rather than on the raw posterior probability to avoid numerical issues, hence we have $\log(p'(\texttt{<st>}|\mat{X})) = \log(\lambda) + \log(p(\texttt{<st>}|\mat{X}))$.

\vspacereduce
\subsection{SCD evaluation metrics}

\begin{figure}[t!]
    \begin{center}
    	\includegraphics[trim={2.3in 1.65in 2.1in 1.7in},clip,width=3.49in]{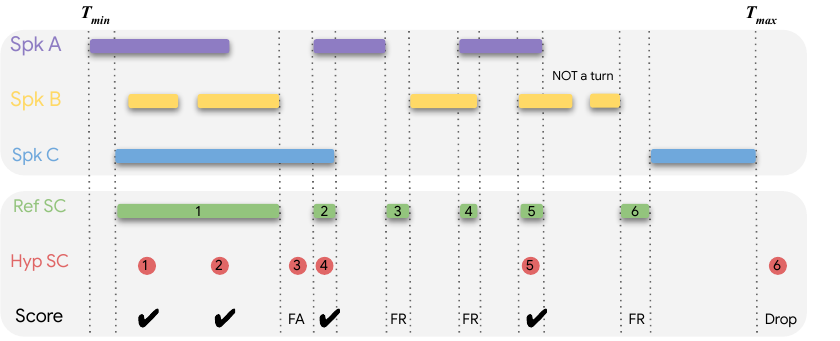}
    	\caption{Illustration of the SCD scoring mechanism for computing the precision, recall, and F1. ``Spk A-C" stands for speaker annotations on a conversational utterance. ``Ref SC" is the reference speaker change intervals. ``Hyp SC" is the predicted speaker change. ``Score" shows the scoring decision of each prediction and reference.}
    	\label{fig:metrics}
	\end{center}
\end{figure}

For SCD evaluation, we compute precision (percentage of model predictions that are true speaker changes), recall (percentage of ground-truth speaker changes that are correctly predicted by the model), and F1 score (the harmonic mean of the precision and recall). We treat the F1 score as a more comprehensive quality metric than the precision/recall rate alone. To compute these metrics, we align predicted and ground-truth speaker changes based on their timestamps, i.e., correct predictions should overlap with the ground-truth labels. Please refer to Fig. \ref{fig:metrics} for an example. For a detailed description of these metrics, please see Sec. 3 of \cite{zhao2023scd}.

\vspacereduce
\section{Experimental setup}
\label{sec:exps}

\vspacereduce
\subsection{Data}

We use various supervised and unsupervised short/long-form data across model training and evaluation. All internal datasets are collected according to Google's Privacy Principles \cite{privacyprinciples} and abide by Google AI Principles \cite{aiprinciples}.

\vspacereduce
\subsubsection{Training}

\textbf{YT-56-U}: This dataset is built by first randomly collecting three million hours of audio from ``speech-heavy" user-uploaded YouTube videos, filtered by user-provided language tags. The three million hours of audio is then further segmented by a Voice Activity Detection (VAD) model and non-speech segments are removed. This yields approximately one million hours of unlabeled audio data. Later, we use a language identification model to select data that corresponds to 56 languages from that unlabelled audio data. \textit{We use this dataset to pretrain the BEST-RQ model}.

\textbf{VS-SUP}: We use a Voice Search dataset consisting of 85 language locales to pretrain the ASR model. There are a total of 1.2 billion short utterances (average duration 4 seconds) from Voice Search traffic. The data is anonymized and human transcribed. \textit{No} speaker change information is available for this dataset. \textit{We use this dataset for ASR pretraining} because short-form supervised ASR training data is significantly larger in volume than long-form data with speaker change labels.

\textbf{YT-SUP}: This is a dataset with audio from YouTube videos that has text transcripts and speaker change labels from 96 languages. We group consecutive segments into a longer unit similar to \cite{lu2021input}. The maximum sequence length for training is 30 seconds. The total quantity of training data is 108k hours, ranging from three hours (Paraguayan Guarani) to 4k hours (Brazilian Portuguese) across locales. \textit{We use this dataset to fine-tune the USM-SCD model}.

\vspacereduce
\subsubsection{Evaluation}

\textbf{YT-96-Eval}: For all languages, we have in total 1,400 hours of internal YouTube long-form evaluation data (no overlap with \textbf{YT-56-U} or \textbf{YT-SUP}) annotated with text transcriptions and speaker changes. On average, we have 15.2h (std: 4.5h) of evaluation data per language and 5 speaker changes per minute of audio in this test set.

\begin{table}
\centering
\caption{Statistics of additional internal and public En-US test sets.}
\resizebox{3.2in}{!}{%
\begin{tabular}{ccccc}
\toprule
\multirow{2}{*}{Testset} & \multirow{2}{*}{Domain} & \multirow{2}{*}{Dur. (h)} & \multicolumn{2}{c}{Average} \\ \cmidrule(lr){4-5}
                         &                         &                           & Turns/min     & Duration/Utt. (min)\\
                         \midrule
AMI \cite{carletta2005ami}                      & Meeting                 & 9.1                       & 10            & 34          \\
Callhome \cite{canavan1997callhome}                 & Telephone               & 1.7                       & 19            & 5           \\
DIHARD1 \cite{ryant2018first}                 & Mixed                   & 16.2                      & 12            & 9           \\
Fisher \cite{cieri2004fisher}                  & Telephone               & 28.7                      & 13            & 10          \\
ICSI \cite{kaldiicsi}                    & Meeting                 & 2.8                       & 13            & 55          \\
Inbound                  & Telephone               & 21.0                      & 9             & 5           \\
Outbound                 & Telephone               & 45.6                      & 13            & 6           \\
\bottomrule
\end{tabular}%
}
\label{table:en_us_eval_data}
\end{table}

\textbf{En-US-Eval}: For American English (En-US), we have additional internal and public test sets, see Table \ref{table:en_us_eval_data}. For the first DIHARD challenge evaluation subset (DIHARD1), we remove all YouTube-derived utterances to avoid evaluating on utterances that might have appeared during training. For Fisher, we randomly sample a subset of 172 utterances for testing\footnoteurl{https://github.com/google/speaker-id/blob/master/publications/ScdLoss/eval/fisher.txt}. ``Outbound" and ``Inbound" are vendor-provided call center telephone conversations between call center operators and customers, initiated by the call center and by customers, respectively. ``Outbound" and ``Inbound" were previously used in \cite{xia2022turn,wang2022highly,zhao2023scd}.

\vspacereduce
\subsection{Modeling details}

We extract 128-dim log-mel filter-bank energies from a 32ms window with a 10ms frame shift as the raw input feature to the model. We use a WordPiece model that has a vocabulary size of 16,384.

The feature encoder contains two 2D-convolution layers of shape $3\times3\times1\times128$ and $3\times3\times128\times32$ (time$\times$frequency$\times$input-channel$\times$output-channel), respectively. The stride size of both convolution layers is 2 on both the time and frequency dimensions. The feature encoder increases the frame rate by 4-fold (i.e., down-sample the frames by 4-fold), from 10ms to 40ms, resulting in a 1,024-dim feature vector. The multi-headed self-attention in the Conformer layers has 8 attention heads. The chunk-wise attention in the Conformer encoder has an 8s context. The convolution kernel size is 5. We run experiments on a model with 1.84 billion parameters, where we have 32 Conformer layers and each layer has 1,536 dimensions.

We use the Adafactor optimizer \cite{shazeer2018adafactor} with a transformer learning rate schedule. For fine-tuning tasks, we optimize the encoder and decoder with separate optimizers and learning rate schedules given that the encoder alone has been pretrained. For the encoder, we use a peak learning rate $3\times10^{-4}$ with 6k warm-up steps, while for the decoder projection layer we use a peak learning rate $5\times10^{-4}$ and 2k warm-up steps. Training was done with a global batch size of 4,096 on TPUs \cite{jouppi2017datacenter}. We monitor the training process on a held-out development set. For all models, we train them for around 40k steps. Empirically \cite{li2022massively}, fine-tuning from a well-trained foundation model only requires a small fraction of training steps compared with training from scratch. In this study, we observe that the model can converge to a reasonable state with as few as 5k training steps, which takes about 6.5 hours of training time with the aforementioned setup.

\vspacereduce
\section{Results}
\label{sec:results}

We compute the WER (for ASR) and SCD precision, recall, and F1 rates as quality metrics. For all evaluations, unless otherwise specified, we use greedy search and aggregate the evaluation data from all 96 languages (\textbf{YT-96-Eval}) to compute the final scores. For WER, we remove speaker change tokens from the scoring.

\vspacereduce
\subsection{Overall system comparisons on YT-96-Eval}
\label{sec:res:overall}

\begin{table}
\centering
\caption{Overall system comparisons on \textbf{YT-96-Eval}. The \textit{w/ SCD} systems are fine-tuned from the corresponding pretrained models with speaker change tokens in the training target; the \textit{w/o SCD} system is trained to perform only ASR.}
\resizebox{3.4in}{!}{%
\begin{tabular}{cccccc}
\toprule
& & BEST-RQ Pretrain & ASR Pretrain & ASR Pretrain & Whisper \\
& & w/ SCD & w/ SCD & w/o SCD & large-v2 \\
\midrule
\multirow{3}{*}{WER} & En-US & 17.1 & \textbf{12.6} & \textbf{12.6} & 16.2 \\
 & 21-lang. & 21.1 & \textbf{16.6} & \textbf{16.6} & 30.1 \\
 & 96-lang. & 34.3 & 30.1 & \textbf{28.8} & - \\
\cmidrule(lr){1-1} \cmidrule(lr){2-2} \cmidrule(lr){3-6}
\multirow{3}{*}{SCD} & Precision & 80.0 & \textbf{82.4} & - & - \\
                     & Recall & \textbf{52.6} & 51.9 & - & - \\
                     & F1 & 63.5 & \textbf{63.7} & - & - \\
\bottomrule
\end{tabular}%
}
\label{table:pre_trained_model}
\end{table}

\begin{table*}[t!]
\renewcommand\thetable{4}
\centering
\caption{En-US results based on \textbf{En-US-Eval}. DIHARD1 and In/Outbound do not have ground-truth text transcripts. The last column shows the evaluation metrics computed by pooling all test sets together.}
\label{table:en_us_detailed}
\resizebox{5.5in}{!}{%
\begin{tabular}{cccccccccc}
\toprule
Metrics & System & AMI & CallHome & DIHARD1 & Fisher & ICSI & Inbound & Outbound & \textit{Pooled data} \\
\midrule
\multirow{2}{*}{WER}       & SCD loss & 39.8 & 33.0 & - & 30.6 & 46.1 & - & - & 33.5 \\
                           & USM SCD & 25.7 & 18.6 & - & 18.4 & 31.5 & - & - & \textbf{20.7} \\
\cmidrule(lr){1-1} \cmidrule(lr){2-2} \cmidrule(lr){3-9} \cmidrule(lr){10-10}
\multirow{2}{*}{Precision} & SCD loss   & 79.4 & 82.0 & 78.8 & 82.6 & 77.8 & 72.8 & 75.1 & 77.6 \\
                           & USM SCD & 91.6 & 84.6 & 92.9 & 94.7 & 90.2 & 94.4 & 91.9 & \textbf{90.8} \\
\cmidrule(lr){1-1} \cmidrule(lr){2-2} \cmidrule(lr){3-9} \cmidrule(lr){10-10}
\multirow{2}{*}{Recall}    & SCD loss & 68.1 & 59.1 & 52.4 & 75.7 & 58.7 & 79.2 & 58.7 & 65.2 \\
                           & USM SCD & 75.3 & 90.8 & 81.7 & 76.5 & 82.7 & 70.1 & 87.3 & \textbf{81.4} \\
\cmidrule(lr){1-1} \cmidrule(lr){2-2} \cmidrule(lr){3-9} \cmidrule(lr){10-10}
\multirow{2}{*}{F1}        & SCD loss & 73.3 & 68.7 & 62.9 & 79.0 & 66.9 & 75.9 & 65.9 & 70.9 \\
                           & USM SCD & 82.6 & 87.6 & 86.9 & 84.6 & 86.3 & 80.5 & 89.5 & \textbf{85.8} \\
\bottomrule
\end{tabular}%
}
\end{table*}

\begin{table}
\renewcommand\thetable{3}
\centering
\caption{Effect of the choice of model parameters to fine-tune. The decoder and input processing layers are always fine-tuned (27M parameters). Evaluated on \textbf{YT-96-Eval}.}
\resizebox{3.2in}{!}{%
\begin{tabular}{cccccc}
\toprule
Fine-tuned & \# Params &  \multirow{2}{*}{WER} & \multirow{2}{*}{Precision} & \multirow{2}{*}{Recall} & \multirow{2}{*}{F1} \\  %
Enc. layers & Trained & \\
\midrule
First 4 & 254M & 35.9 & 83.8 & 35.6 & 50.0 \\
Last 4 & 254M & 30.4 & 82.2 & 44.6 & 57.8 \\
First 4 \& last 4 & 480M & \textbf{30.1} & \textbf{84.0} & \textbf{52.5} & \textbf{64.6} \\
All & 1.84B & \textbf{30.1} & 82.4 & 51.9 & 63.7 \\
\bottomrule
\end{tabular}%
}
\label{table:subcomponent}
\end{table}

We first study the choice of the pretrained model. The results are summarized in the first two columns of Table \ref{table:pre_trained_model}. The SCD models fine-tuned from the two pretrained models yield comparable SCD F1 scores (0.3\% relative difference), suggesting that they are comparable in terms of detecting speaker change events. The SCD model fine-tuned from the ASR model has significantly better WER (30.1 vs 34.3 across 96 languages; a 12.2\% relative reduction), demonstrating the benefit of ASR-pretraining on the word-level SCD task.

Next, we study the trade off between ASR and SCD. We fine-tune from the ASR-pretrained checkpoint to construct \textit{ASR Pretrain w/o SCD} that does not have the speaker change token in the training target, resulting in a WER of 28.8\%. Therefore, with the proposed approach, adding the SCD capability to the ASR model would result in a 4.5\% relative WER regression.

To provide additional context, we compare the WER of the USM-SCD model with a strong publicly available ASR model Whisper \cite{radford2023robust} (large-v2, 1.55B parameters) that was trained on more than 400k hours of transcribed ASR data. We select 21 top performing languages from Whisper (which achieve WER lower than 40\% on \textbf{YT-96-Eval}), and the results are shown in the last column of Table \ref{table:pre_trained_model}. We observe that although adding the SCD capability to an ASR model hurts the WER, the resulting USM-SCD model still has a better ASR performance on YouTube data compared to Whisper.

\vspacereduce
\subsection{Effect of sub-components to fine-tune}
\label{sec:res:sublayers}

We now study which model parameters to fine-tune. For this experiment, we always fine-tune from the ASR pretrained model given the results in Sec. \ref{sec:res:overall}. Given that we are using a different data source (i.e., \textbf{YT-SUP}) for SCD training and we modify the training targets, we always fine-tune the \textit{feature} encoder, input projection, and decoder projection layers, which consist of 27M trainable parameters. A preliminary experiment suggests that \textit{only} fine-tuning the feature encoder, input projection, and decoder projection layers does not converge well. Therefore, we selectively fine-tune certain layers of the Conformer encoder and freeze the rest of the parameters. All models are trained for 40k steps. The results are in Table \ref{table:subcomponent}. We observe that optimizing the last 4 layers is significantly better than optimizing the first 4 layers both in terms of WER and SCD metrics. Interestingly, optimizing both the first 4 and last 4 layers (i.e., 8 of 32 layers) gives the best ASR and SCD performance, which only accounts for $\sim$26\% of the trainable parameters.

\vspacereduce
\subsection{Effect of the speaker change token posterior scaling}

\begin{figure}
\centering
\includegraphics[trim={0.2in 0.2in 0.2in 0.2in},clip,width=3in]{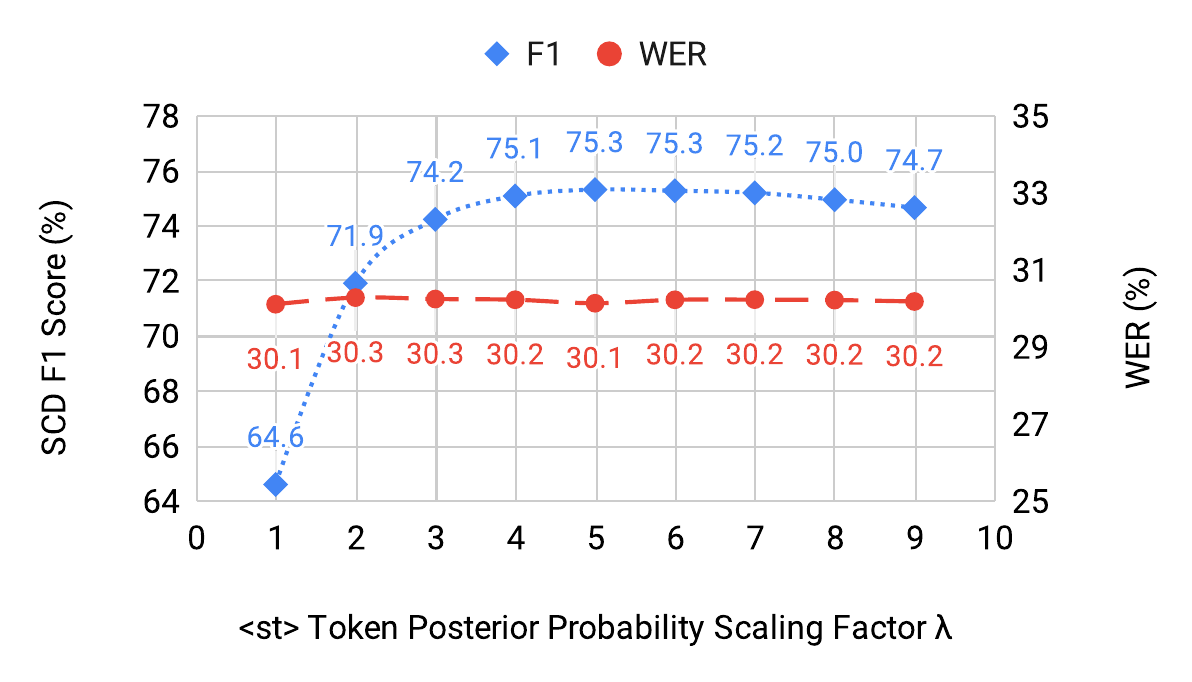}
\caption{SCD token \texttt{<st>} posterior probability scaling results on \textbf{YT-96-Eval}.}
\label{figure:prior_scaling}
\end{figure}

Next, we study the effect of the speaker change token posterior scaling factor (cf. Sec. \ref{sec:method:scd_posterior}). Based on the results in Sec. \ref{sec:res:sublayers}, we use the model that is only fine-tuned on the first 4 and last 4 layers of the Conformer encoder (480M trainable parameters). We run experiments (see Fig. \ref{figure:prior_scaling}) by setting the factor $\lambda$ from 1.0 to 9.0, with a step size of 1.0. Note that this experiment does not require retraining the model since the posterior scaling happens during inference. We observe that the posterior scaling does not significantly affect the ASR quality, with the maximum WER difference being less than 0.7\% \textit{relative} (i.e., from 30.1\% to 30.3\%). More importantly, the scaling factor brings large gains in terms of SCD quality. Compared with the baseline configuration where there is no SCD posterior scaling (i.e., scaling factor 1.0), the best posterior scaling factor of 5.0 increases the SCD F1 score from 64.6\% to 75.3\%, a 16.6\% relative improvement.

\vspacereduce
\subsection{En-US quality analysis}

For En-US, there are additional internal and public datasets that have speaker change labels (Table \ref{table:en_us_eval_data}). We evaluate the \textit{USM-SCD} model fine-tuned from ASR on these datasets. We only fine-tune the first 4 and last 4 Conformer encoder layers, and the SCD posterior scaling factor is set to 5.0 during inference. The per-testset results are summarized in Table \ref{table:en_us_detailed}. We also include the best performing system from \cite{zhao2023scd} (denoted as \textit{SCD loss}, 27M parameters monolingual En-US model) as a comparison. The \textit{SCD loss} system is trained with an SCD-optimized training loss on a super-set of the En-US portion of \textbf{YT-SUP}, with 2k hours of additional training data from other domains. We observe that the \textit{USM-SCD} system performs much better than the \textit{SCD loss} system, achieving 21\% relative F1 score improvement. The precision and recall rates increase by 17.0\% and 24.8\% relative, respectively.

\vspacereduce
\section{Discussion and Conclusion}
\label{sec:conclusion}

In this work we propose a multilingual SCD model that supports 96 languages. We take advantage of recent advances in large speech foundation models to construct this USM-SCD model and study its properties through a series of ablation studies. We find that ASR-pretraining is crucial to model performance. We observe that we only need to fine-tune roughly one-quarter of the trainable parameters to achieve the best overall performance compared to fine-tuning all parameters. We also show that an inference-time SCD token posterior scaling that requires no additional computation can result in a 16.6\% relative improvement in the SCD F1 score. Finally, compared with our previous monolingual En-US SCD model, the USM-SCD model outperforms it by 21\% in terms of SCD F1 score. Based on benchmarks on TPU v5e \cite{cloudtpuv5e}, the USM-SCD model can run inference at 60x faster than real-time (batch size 1), demonstrating the application potential of this model. Possible future directions include replacing the CTC architecture with a fast RNN-T implementation and applying the token-level training loss proposed in \cite{zhao2023scd} to further boost model quality. It is also interesting to explore multi-output RNN-T joint networks \cite{wang2023multi} to decouple the ASR and SCD tasks.

\vspacereduce
\section{Acknowledgements}
The authors would like to thank Wei Han for the Whisper model evaluation setup, and Olivier Siohan, Parisa Haghani, Ignacio Lopez Moreno, and Pedro Moreno Mengibar for reviewing this work.

\bibliographystyle{style/IEEEbib}
\vspacereduce
\newpage
\bibliography{references}

\end{document}